\documentclass{PoS}
\newcommand{\be}{\begin{eqnarray}}
\newcommand{\ee}{\end{eqnarray}}
\title{Gauge topology and confinement: an update}

\ShortTitle{Gauge topology}

\author{\speaker{Edward Shuryak}\\
        Department of Physics and Astronomy, Stony Brook University, Stony Brook NY 11794 USA\\
        E-mail: \email{edward.shuryak@stonybrook.edu}}

\abstract{In the instanton ensemble the  fermionic zero modes
collectivize and break chiral symmetry. Recent studies of resulting zero
mode zone confirm its very small width and overall importance
for lattice simulations.
 Confinement however has been
related with completely different topological objects, the magnetic monopoles. Instanton
constituents -- instanton dyons, discovered at nonzero holonomy by   
Pierre van Baal and others -- are able to explain both confinement and chiral symmetry breaking. The talk
summarizes recent works deriving the instanton-dyon mutual interactions,
and statistical studies of their ensemble. At high density the screening
is robust enough to do it analytically, in the mean-field-type approach:
we call this limit Dense Dyonic Plasma (DDP). Above $T_c$ 
 the classical interaction between the dyons induce strong correlations and should be studied by
 by direct numerical simulations. Those works are now in progress.}

\FullConference{9th International Workshop on Critical Point and Onset of Deconfinement - CPOD2014,\\
		17-21 November 2014\\
		ZiF (Center of Interdisciplinary Research), University of Bielefeld, Germany}

\begin{document}

\section{Instantons,   the Zero Mode Zone, and chiral symmetry breakings }
Instantons  \cite{Belavin:1975fg}  are Euclidean 4-dimensional topological solitons of the Yang-Mills gauge fields. 
While small-size instantons have large action and exponentially suppressed,
instantons of the size $\rho\sim 1/3 \, fm$ have substantial density $n\sim 1\, fm^{-3} $\cite{Shuryak:1981ff}
and are therefore a very important ingredient of the
gauge fields in the QCD vacuum.

Chiral anomalies induce fermionic zero modes of instantons, which generate the so called 't Hooft interaction between $2N_f$ fermions, which 
 explicitly violates the $U_A(1)$ chiral symmetry. 
 Statistical mechanics of instanton ensemble,  including  't Hooft interaction
 to all orders,  known as  
 the Interacting Instanton Liquid Model,  has been solved numerically, by brute force, two decades ago,  for a review  see
\cite{Schafer:1996wv}. Not going into details, we remind the
 central concept crucial for understanding of the spontaneous $SU(N_f)$ chiral symmetry breaking: 
{\em a  collectivization} of these fermionic zero modes
into the so called 
{\em Zero Mode Zone} (ZMZ) of quasi-zero Dirac eigenstates. When those  have
a finite density at zero eigenvalues $n(\lambda \rightarrow 0)\neq 0$, quark condensate appears.
What was known for decades 
 (and few times rediscovered by lattice practitioners) is that the diluteness of topological ensembles
 leads to small  ``quark hopping" matrix elements of the Dirac operator, between  zero modes of different solitons (i,j). In the instanton liquid with 
 4-density $n$ and size $\rho$ they are of the order of  \be \sigma \sim |<i|D|j>| \sim \rho^2 n^{1/2} \sim 20\, MeV\ee  
 and so is the {\em characteristic width} of the ZMZ
\footnote{ Incidentally, it  is of the order of quark masses used on the lattice in the previous decades: therefore
 simple linear/log expressions of chiral perturbation theory   did not work well.}.
 While the ZMZ states, with Dirac eigenvalues $|\lambda |<\sigma \sim 20\, MeV$, make
  only tiny ($\sim 10^{-4}$) subset of all fermionic states on the lattice,   they are responsible
 for a significant fraction of most hadronic masses.  Studies of ZMZ
 states, in the context of the instanton liquid and lattice, has been made in 1990's.
  Recently  Graz group  \cite{Glozman:2012fj}  had, once again, 
 demonstrated its importance by  {\em removing} all the ZMZ states from the fermionic propagators: they observe
 a completely different hadronic spectrum, in which 
  both of the $U_A(1)$ and  the  $SU(N_f)$  chiral symmetry breaking is gone.
 So, gauge topology is not just some
 cute little effect which some theory-inclined people tend to study: it is in fact responsible
 for a significant fraction of hadronic masses (and thus our own weight)!

Another feature observed by  the Graz group is strong reduction of the statistical
fluctuations in the ensemble, which is of practical importantance for lattice community.  In spite of many millions of lattice variables  it is hardly surprising, since
there are only about a dozen or so  instantons in the box. Let me predict that current efforts  to improve efficiency of
lattice updates using   multi-grid algorithms for small Dirac eigenvalues  
are bound to eventually rediscover the instanton ensemble, with a delay of 3 decades or so. 

At the end of this introductory section, let me mention progress in application of instantons in other directions. 
Instanton-induced t' Hooft interaction generates attraction between quarks, well explaining the ``scalar diquark"
 phenomenon well known in hadronic phenomenology.  Scalar diquarks become Cooper pairs of  color superconductivity at high density \cite{Rapp:1997zu}. Instantons also generate important spin effects in hadronic reactions, such as spin
 asymmetries \cite{SPIN}. 

 On a more theoretical note, let me remind
 that   the  Seiberg-Witten solution of $\cal{N}$=2 supersymmetric gluodynamics, based on the celebrated elliptic curve,
 was derived by the explicit calculation of all instanton amplitudes  \cite{Nekrasov:2002qd}. Recently \cite{Basar:2015xna} it has been further related to
 instantons in some quantum-mechanical systems. 
 
 There is also recent progress in the latter field, fueled by exact global boundary condition relating
  ordinary perturbation theory and those on top of the instantons, pioneered by Zinn-Justin.  Dunne and Unsal \cite{Dunne:2014bca}
have suggested even more direct relation between them: however its meaning
remains  unclear.
Our modest contribution to that development was
 explicit   calculation of the three-loop corrections to the instanton density \cite{Escobar-Ruiz:2015nsa} by  the  direct evaluation of all Feynman
 diagrams: the result is consistent with the relations mentioned.

\section{The instanton-dyons and confinement}

Semiclassical theory of instantons is parametrically good at high $T$, as the effective coupling gets there weak:
but they are there strongly suppressed.  When light fermions are present in the theory, at high $T$
the ``instanton liquid" is absent, and 
instantons can exist (i) either in form of neutral instanton-antiinstanton molecules \cite{Ilgenfritz:1988dh}, or
(ii) single instantons, with the density proportional to all quark masses $\Pi_f m_f$, as argued by
't Hooft in his original paper on fermion zero modes. Since even the unphysical quark masses used
in lattice simulations are still rather small, only recently the Bielefeld groups \cite{SHARMA} 
had clearly identified a contribution in the Dirac eigenvalue density of the kind $n(\lambda)\sim m^2 \delta(\lambda)$,  corresponding to
single instantons.

  Decreasing the temperature below  $2T_c$
one  finds 
a nontrivial average value of the Polyakov line $<P>\neq 1$ , indicating that an expectation value of the gauge potential  
is nonzero $<A_4>=v\neq 0$. This calls for re-defining the boundary condition of $A_4$ at infinity, for all solitons
including instantons. That  lead to
1998 discovery \cite{Kraan:1998sn,Lee:1998bb} that nonzero $v$
  splits
instantons  into $N_c$ (number of colors) constituents, the selfdual {\em instanton-dyons}\footnote{
They are  called  ``instanton-monopoles" by Unsal et al, and are similar but not identical
to ``instanton quarks" discussed by Zhitnitsky et al.} .
Since these objects have nonzero electric and magnetic charges and source
Abelian (diagonal) massless gluons, the corresponding ensemble is 
an ``instanton-dyon plasma", with long-range Coulomb-like forces between constituents.  
By tradition the selfdual ones are called $M$ with charges $(e,m)=(+,+)$ and $L$ with charges $(e,m)=(-,-)$, the anti-selfdual antidyons are called  
 $\bar{M}$, $(e,m)=(+,-)$ and  $\bar{L}$, $(e,m)=(-,+)$.

Diakonov and collaborators (for review see \cite{Diakonov:2009ln} )
 emphasized that, unlike the (topologically protected) instantons, the dyons interact directly with
 the holonomy field. They suggested that since such dyon (anti-dyon)  become denser
at low temperature, their back reaction  may overcome perturbative holonomy potential and drive it
to its confining value, leading to  vanishing of the mean Polyakov line, or confinement.

Specifically, Diakonov and collaborators focused on the self-dual sector $L,M$ and studied the one-loop
contribution to the partition function \cite{Diakonov:2004jn}. The volume element of the moduli space was
written in terms of dyons coordinates as a determinant of certain matrix $G$, to be referred to as Diakonov determinant. In a dilute limit it leads to
 Coulomb interactions between the dyons, but in the dense region it becomes strongly repulsive, till at certain density
the moduli volume vanishes. 

  Interesting  semi-classical description
 of the confining regime has been found by 
 Unsal and collaborators~\cite{UNSAL}   in a carefully devised setting of softly broken supersymmetric models.
 While the setting includes a compactification on a small circle, with  weak coupling and
 an exponentially  $small$ density of dyons, the minimum at the confining holonomy
  value is induced by the repulsive interaction in the dyon-antidyon molecules (called  
 $bions$ by these authors). 
 The crucial role of the supersymmetry is the cancellation of the perturbative Gross-Pisarski-Yaffe-Weiss (GPYW)  holonomy potential:
 as a result, in this setting there is no deconfined phase with trivial holonomy at all, unless supersymmetry is softly broken.
  Sulejmanpasic and myself \cite{Shuryak:2013tka} proposed a simple analytic model for the dyon ensemble
with  dyon-antidyon ``repulsive cores", and have shown how they may naturally
 induce confinement in dense enough dyonic ensemble.

Traditional explanation of the confinement is often given in terms of particle-monopoles and 
dual superconductor model of 't Hooft and Mandelstamm: so one may ask about 
any relation between particle-monopoles and
instanton-dyons. 
In $\cal{N}$=2 SYM (Seiberg-Witten theory) both are under theoretical control, 
so one can calculate how each of them contribute to partition function. Remarcably \cite{Poppitz:2011wy}
 one finds that the results  are equal! One can see that  particle-monopoles provide better
behaved sum at low-$T$, while the instanton-dyons better converge at the high-$T$ end,
but both describe the same physics. Unfortunately, in non-supersymmetric theories,
without such explicit expressions available, the relation between the two remains unclear.

\section{Classical dyon-antidyon interaction and the streamlines}

As it is well known, the selfduality of instantons or their constituents, combined with the famous Bogomolny
inequality, relates the classical action to the topological charge and thus protect them from classical
interactions. Any number of $M,L$ dyons, or $\bar{ M},\bar{ L}$  have the same total action, independent on their positions:
as a result one has exact moduli spaces. (The interaction only appears at the one-loop order,
as we already mentioned above.)

This  however is no longer true when one mixes together selfdual and anti-selfdual objects \footnote{A
famous way to avoid it is to use supersymmetry-based holomorphy arguments, which forbid  interaction between the sectors.
 This nice escape root has been taken by many, but for me dealing
 with only non-existing toy theories seemed to be too  heavy price to pay for mathematical convenience. }.
Instanton-antiinstanton
 configurations can be mapped via the so called  {\em  streamline}, a one-parameter set of solutions defined by a condition
 that the driving force (current), while nonzero, is tangent to the set. 
   A practical way to generate them is to follow the {\em gradient flow}, starting from some initial ansatz,
   as was done numerically  for the instanton-antiinstanton in the double-well potential in \cite{Shuryak:1987tr}. 
 
  Let me add a new comment on quantum-mechanical streamline, related with recent renewed interest
  in ``resurgent series". Not going into formulae here, let me just remind that it is based on ``three whales", and the
  coefficients have
  three indices, corresponding to three basic elements: (i) $g$, the coupling constant,
  (ii) tunneling series over powers of $exp(-const/g)$ where the constant is defined by instantons, and (iii) the powers of $log(1/g)$.  The first are well known perturbative series, the second is the instanton series: but what is the ``third whale"?
   These terms originate  come from the ``lower part" of the instanton-antiinstanton  streamlines,
  or from the ``$I\bar{I}$ molecules". This streamline goes all the way to trivial configuration, with zero field and action,
  $S(\Delta t \rightarrow 0)=0$. Yet the semiclassical theory is only valid when the action 
   remains large
   $S (\Delta t)\sim exp(-t)/g\gg 1$, which creates nontrivial lower integration limit on $\Delta t> log(1/g)$ ,
   and thus contribution to the total energy. In gauge theories  
   such terms, coming from the repulsive core in the instanton-antiinstanton channels, are also known,
   and in fact were the important element of the Interacting Instanton Liquid Model.

For gauge field instantons  the first study of the instanton-antiinstanton
streamline was done by Balitsky and Yung \cite{Balitsky:1986qn}, in the
large-distance approximation.  Since classical 
 instanton-antiinstanton problem is intrinsically conformal, one can perform a conformal transformation into
a co-central configuration, relating the gauge theory  and the double-well instantons.
Using this method  Verbaarschot  \cite{Verbaarschot:1991sq}  found a quite accurate analytic
approximation to the instanton-antiinstanton streamline.

\begin{figure}[h!]
\includegraphics[scale=0.6]{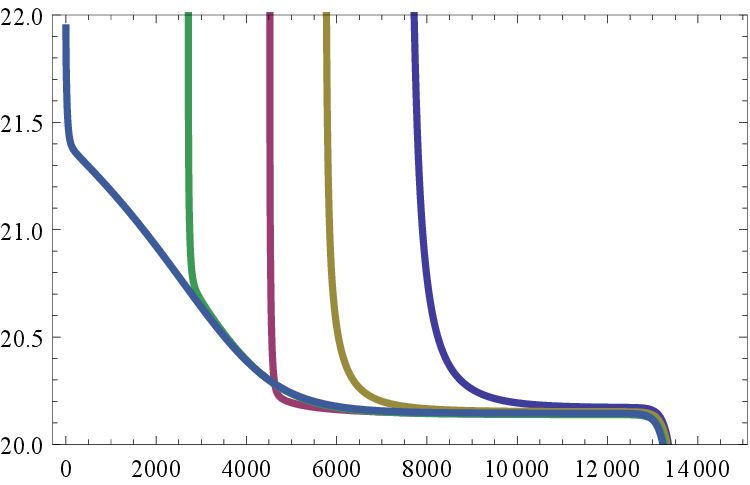}
\includegraphics[width=6.5cm]{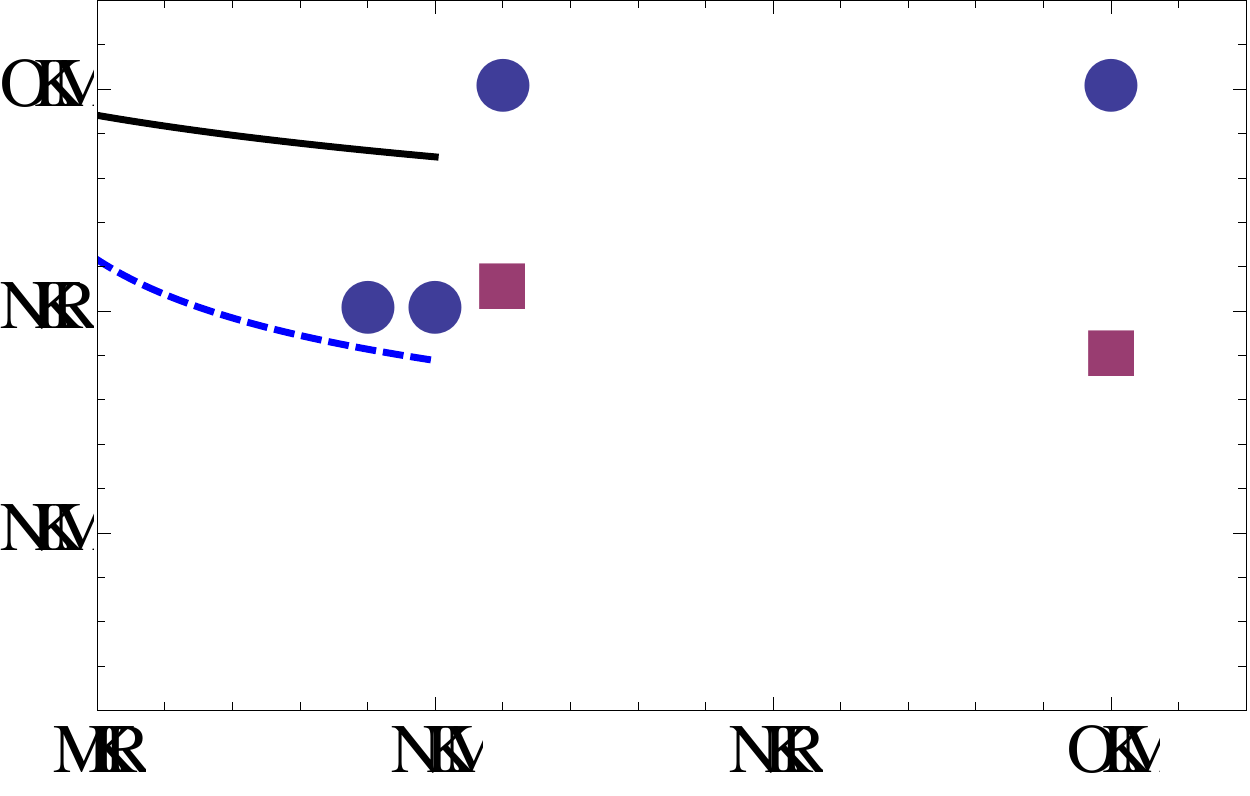}
\caption{
(a) 3d-action $S_3$ for the $M\bar{M}$ dyon pair, as a function of computer time of the gradient flow.
Subsequent lines are for  separations $|r_M-r_{\bar{M}} |v=0$, $2.5$, $5$, $7.5$, $10$,  from right to left in the graph. The action of two well separated dyons is 23.88.
(b) The electric $M_E/T$ (dashed line) and magnetic $M_M/T$ (solid line) screening masses in  versus $T/T_c$.
The points are SU(2) lattice data from shown for comparison, (blue) circles are  $M_E/T$, (red) squares are $M_M/T$
}
\label{Fig1}
\end{figure}

Larsen and myself have recently calculated the dyon-antidyon $M\bar{M}, L\bar{L}$
 streamline \cite{Larsen:2014yya},  by performing gradient flow  
calculations on a 3d lattice. We start with two separated and ``well combed" dyons,
with $A_4$ going into the same direction at all infinities, and then start to minimize the action
along the gradient flow direction.
Many details can be found in the original paper, the   key Fig. \ref{Fig1}(a)
shows how the action depends on the computer time.
 The gradient flow process was found to proceed via the following stages: \\
 (i) {\em near initiation}: starting from relatively arbitrary ansatz one finds rapid disappearance of artifacts and convergence toward the streamline set\\
 (ii) following the {\em streamline itself}. The action decrease at this stage is small and steady. The dyons basically approach each other, with relatively small deformations:
 thus the concept of an interaction potential between them makes sense at this stage\\
 (iii) a {\em metastable state} at the streamline's end: the action remains constant, evolution is very slow and consists of internal deformation of the dyons rather than further approach\\
   (iv) {\em rapid collapse} into the perturbative fields plus some (pure gauge) remnants\\   
 
The existence of the stage (iii)  has not been anticipated. All configurations corresponding to it
have the same action, and -- within our accuracy the same dyon-antidyon distance. 
One can perhaps lump all of them into a new metastable configuration, perhaps  identifiable on the lattice.

Our other comment is that even at the end of the streamline the  action value is not that far from the sum of the two dyon masses. In other words, the
classical interaction potential appears to be in a sense numerically small, although parametrically
dominant:  a welcoming feature for statistical mechanics simulations.  

Last but not least, we do observe the $universality$ of the streamline. It means that, independently on the 
initial dyon separation and the details of the initial states,  the gradient flow path proceeds
through essentially the same set of configurations, at stages (ii-iv). Thus one-parameter characterization of those is possible. A parameter 
we found most practical in this work is simply its $lifetime$ -- duration in our computer time $\tau$, needed for a particular configuration to reach a  final collapse.
, following through the so called ``streamline" set of configurations,
till their collapse to perturbative fields.

Finally, let me add that 
the last $M \bar{L}, L \bar{M}$  channels were also investigated  recently.
(It was not done in the original paper because in this case the configurations are time-dependent
and thus 3d lattice was insufficient.) The interaction is repulsive, approximately sign flipped version
of that discussed above.

\section{Dense dyonic plasma}

In this short summary there is no place to discuss numerical parameterizations and other details:
so we only explain the key issues of the problem. 
The classical dyon-antidyon interaction is about an order of magnitude stronger than the one-loop
effects studied before. The potential  (normalized so that partition function includes $exp(-V)$) 
is $V\sim O(1)$ and therefore in general it is strong enough to make  
the dyon plasma to be strongly coupled. Such potential in general generates strong correlations, requiring brute force methods to study them. Those will be discussed at the end of this talk.

However, if the dyon plasma is sufficiently dense, it also generates strong
 screening effects. Those effectively weaken the 
interaction, till above certain density it become manageable via  the usual Debye-Huckel approximation. 
For brevity, we call this regime ``dense dyon plasma", DDP for short.
Its theory has been developed in the paper by Y.~Liu,  I.~Zahed, and myself \cite{FIRST} . 
The dyonic plasma is dense enough at low temperature. Furthermore, as we found 
that the holonomy needs to be at the confining value $v=\pi T$, the applicability of this work
is restricted to only the confining phase $T<T_c$.

While the one-loop effects are subleading in a pairwise interactions at large distances, it turns out that 
manybody effects of the moduli volume gets strongly repulsive at higher density. Inclusion of the one-loop
effect via Diakonov determinant is also included: after some changes of variables we introduce new
bosonic fields for both classical and one-loop Coulomb-like forces. The formulae are to be followed in
the original paper, as they are quite involved. We found that running coupling
induces subtle changes in the interaction: the effective coupling (given by the interaction formfactor
at most important momentum) changes sign, which manifests itself directly in crossing of the
electric and magnetic screening masses. The results are shown 
in Fig.~\ref{Fig1}(b)
we display the results  for $M_{E,M}/T$ in the range $(0.5-1)\,T_c$ 
versus $T/T_c$. 

The points at $T>T_c$ are lattice data shown for comparison. 
Note that the electric mass drops down sharply at $T_c$, and, as a result,  in the region we study
magnetic screening is stronger  $M_M>M_E$, while above $T_c$, in a more familiar
QGP region, the opposite relation is true, $M_M<M_E$. This switching of the magnitude of the two screening masses
is dynamically explained in the paper.

Many more results on the DDP properties has been calculated in this approach: the free energy,
densities of the dyons, string tension, correlators of various Polyakov and Wilson lines, etc.
All of them can and should be compared to lattice studies.

\section{Chiral symmetry breaking and instanton-dyons}

Topological index theorem require one (fundamental) 4d fermionic   zero mode per unit topological charge.
An instanton has $N_c$ constituent dyons: and the issue where the fermion mode is located .
The question has been answered
quickly by van Baal and collaborators: the fermionic zero modes is located primarily at
one of the dyons, jumping to another one as a function of holonomy value and/or periodicity phase of the fermions,
as certain relation between them is fulfilled. 
For the simplest color group $SU(2)$ mostly discussed so far, the physical (anti-periodic) quarks have zero mode on $L$ dyons, while the periodic once have zero modes located at  $M$ dyons. 

 Sulejmanpasic and myself \cite{Shuryak:2012aa} have analyzed a number of  phenomena induced by
 the fermionic zero modes of the instanton-dyons such as the formation of clusters  
 (molecules or bions) at high temperature and the spontaneous breaking of chiral symmetry 
 at low temperature.  Several puzzles originating from lattice observations have been explained on this way.

 For example, quenched simulations of the gauge theory had shown deconfinement at some $T_c$.
 Although no fermions were included in the theory, one can add them as an analysis tool later,
 and check in particular where chiral symmetry is broken. Quite early it was observed that
 at $T<T_c$ the chiral symmetry is broken, and restored above $T_c$, 
 for physical (anti-periodic) fermions. Later it was observed, that using periodic fermions instead
 one gets a very different quark condensate, which does not disappear at deconfinement and persisted well above $T_c$. It was puzzling
 in the instanton framework, since there is one zero mode per instanton, for both periodic and antiperiodic
 fermions.  It became clear in the dyon setting, which allows
 different action -- and thus density -- for $M$ and $L$ dyons. Above $T_c$ the $M$ dyons are ``lighter" than $L$
 ones, and more numerous. Periodic fermions ``see" the ensemble of $M,\bar{M}$, while anti periodic  
``see" the ensemble of $L,\bar{L}$ ones: so the difference is natural. 

 
 The mean field approach to the problem with fermions is discussed in the second paper of the
 dense plasma series, of  Liu,Zahed and myself \cite{SECOND}. After certain manipulations with fermions, we bosonize the
 partition function. The 
  scalar mesonic field $\sigma=\bar q q$ obtains the corresponding gap equation on its VEV,
  only slightlyy more complex than in the Nambu-Jona-Lasinio-type models.
 We explicitly solve it for different number of flavors $N_f$ and discuss how that nonzero solution 
 deviates from previous studies with instantons.
 
\section{Numerical simulations of the dyon ensemble}

The first paper attempting direct Metropolis study of the dyon ensemble was by
Faccioli and  myself \cite{Faccioli:2013ja}.
It has a number of  methodical differences from the instanton ensembles: the interactions are long-range
and thus more sensitive to boundaries. We decided to put the dyons on a 3-d sphere rather than the tori (used
in many other works including lattice gauge theory) with a simplified Coulomb on a sphere.  
Other methodical details should be found in the paper.


The one-loop Diakonov determinant  generates repulsive interaction at high density:
this feature we have studied in detail. However the
 main thrust of that paper  was on fermions and chiral symmetry breaking, as a function of the
number of flavors. Generic phenomenon, familiar already from the instanton liquid, is that
at high $T$ -- low dyon density -- one finds clustering of $L\bar{L}$ dyons, while at high
density the quasi zero modes get collectivized and the chiral symmetry gets broken.
This mechanism is very robust for $N_f=1,2,3$ flavors but falters at $4$ and perhaps does not work
at 5 and more. 

As we already mentioned above, the classical dyon-antidyon interactions are strong enough
to make the ensemble ``strongly coupled", with significant correlations between particles. 
This interaction has not be included by Faccioli and myself: but its presence makes 
direct statistical simulations even more necessary.

New series of simulations is now performed by Larsen and myself \cite{LS_stat}. Using GPU instead of CPU
for evaluation of the determinants (of the Diakonov matrix)   had speed up the program quite a bit.
The main issue we focus on is the back reaction of the dyons on the holonomy value $v$.
We do observe the confinement phase transition, from small $v$ at high $T$ to its confining value
below certain $T_c$. The part which is more delicate and still needs more work is
to self-consist the dyon density and holonomy values. Generalization of this work to
theories with fermions is not yet done but is expected to be rather straightforward.  

\section{Summary}
 After a certain minimum of activity, studies of  gauge topology are getting more active again.
 New round of works on quantum mechanical instantons appeared,
 focusing on trans-series and relations between the perturbative and instanton series. 
A nice  bridge to supersymmetric models has been developed by Unsal and collaborators.  

 The semiclassical theory of gauge topology at nonzero temperatures, including nonzero holonomy
and thus  based on the instanton-dyons, is becoming quantitative.
 Its applicability region cover approximately the temperature range $0.5<T/T_c<2$. At the upper range the ensemble
is a dilute gas made of very correlated clusters. At the lower $T$ the system is  {\em dense dyonic plasma},
 DDP,
with strong effective screening,  allowing to use a Debye-Huckel-like mean field theory. 

Unlike the previous topological models, now both nonperturbative
phenomena -- {\em confinement and chiral symmetry breaking} --  are explained in the same framework.
Confinement is generated by the back reaction of the dyons on the holonomy potential, while the chiral symmetry breaking 
is induced by collectivization of the fermionic zero modes of $some$ dyon type, similar to that in the instanton liquid models.
They both require dense enough ensemble, and at current accuracy we don't separate
 their effective critical temperatures: perhaps those would be revealed by future  more accurate studies.
On the lattice, the situation is similar: there are indications to separation of deconfinement and
chiral phase transitions at higher $N_f$, but the phenomenon is not yet well quantified.

{\bf Acknowledgements}
A lot of credit in this subfield go to our friends who are no longer with us, Pierre van Baal and Dmitri Diakonov.
My progress in understanding those issues would not be possible  without my collaborators on the projects discussed above,  P.Faccioli, I.Zahed, T.Sulejmanpasic and R.Larsen. My  work was supported by the U.S. Department of Energy under Contracts No.
DE-FG-88ER40388.

\end{document}